\documentclass[12pt,dvips]{article}
\usepackage{epsfig}
\def\be{\begin{equation}} 
\def\ee{\end{equation}}
\begin{document}
\begin{figure}[!t]
\centering
  \epsfxsize=13.cm\epsfysize=16cm \epsffile{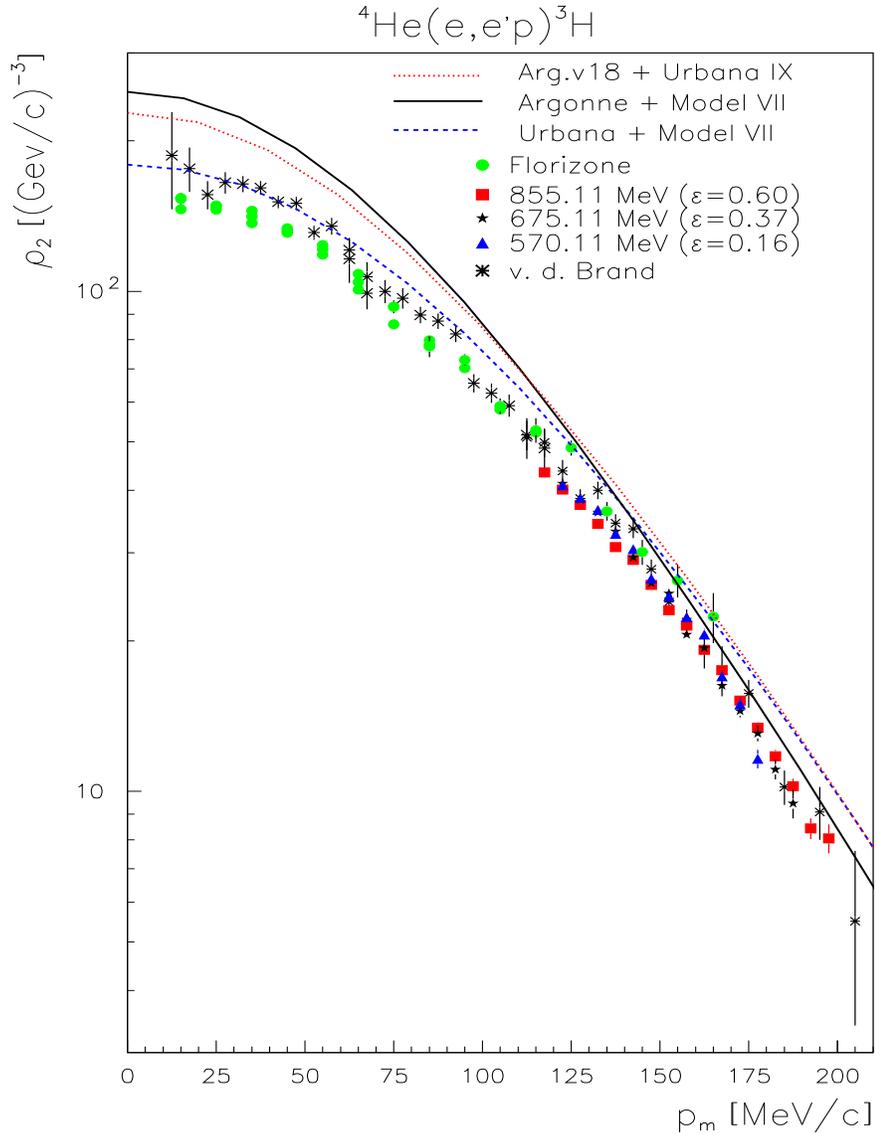}
\caption{\footnotesize{The $p-t$ momentum distributions}}
\label{fig1}
\end{figure}
\begin{figure}[!h]
\centering
  \epsfxsize=15.cm\epsfysize=20cm \epsffile{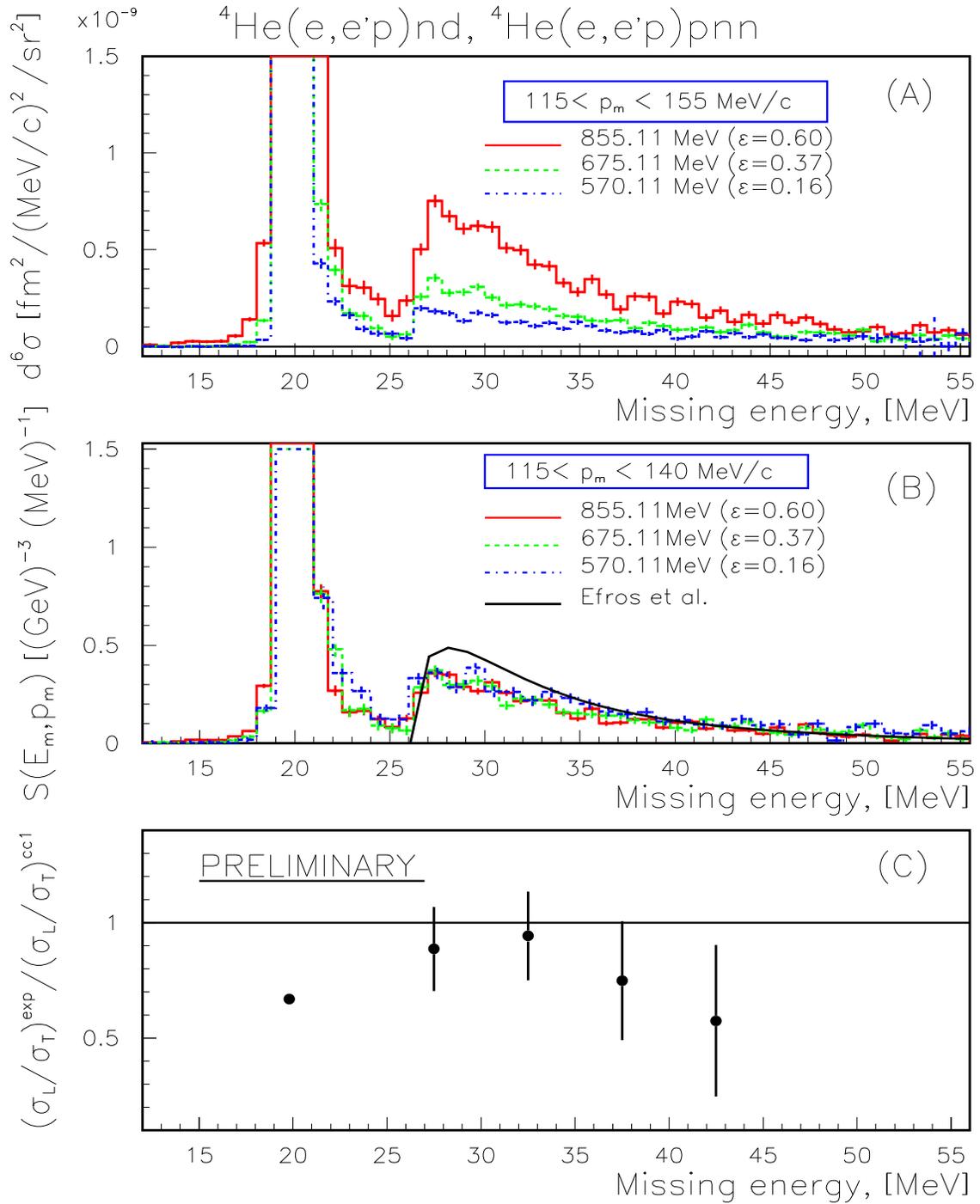}
\caption{\footnotesize{Preliminary results shown as a function of missing
energy. (A) Six-fold differential cross section for the three 
values of the virtual-photon polarization $\epsilon$ in the missing momentum 
range  from 115 to 155 MeV/c. (B) Spectral function. (C) The ratio 
$\sigma_{L}/\sigma_{T}$ (both the 2-body and continuum channels).}}
\label{fig2}
\end{figure}

\begin{center}
{\large{\bf{The Longitudinal and Transverse Response of the $^{4}\!He(e,e'p)$ Reaction in the Dip Region}}}
\end{center} 
\begin{center}
{\small{ A.~Kozlov$^{1}$, K.A.~Aniol$^{5}$, P.~Bartsch$^{2}$, 
D.~Baumann$^{2}$, 
W.~Bertozzi$^{3}$, R.~B\"{o}hm$^{2}$, K.~Bohinc$^{10}$, J.P.~Chen$^{4}$,
D.~Dale$^{6}$, L.~Dennis$^{8}$, S.~Derber$^{2}$, M.~Ding$^{2}$,
M.O.~Distler$^{2}$, A.~Dooley$^{8}$, P.~Dragovitsch$^{8}$, M.B.~Epstein$^{5}$,
 I.~Ewald$^{2}$, 
K.G.~Fissum$^{3}$, R.E.J.~Florizone$^{3}$, J.~Friedrich$^{2}$, 
J.M.~Friedrich$^{2}$, R.~Geiges$^{2}$, S.~Gilad$^{3}$, P.~Jennewein$^{2}$,
M.~Kahrau$^{2}$, M.~Kohl$^{7}$, K.W.~Krygier$^{2}$, A.~Liesenfeld$^{2}$,
D.J.~Margaziotis$^{5}$, H.~Merkel$^{2}$, P.~Merle$^{2}$, U.~M\"{u}ller$^{2}$, 
R.~Neuhausen$^{2}$,
T.~Pospischil$^{2}$, G. Riccardi$^{8}$, R.~Roche$^{8}$, G.~Rosner$^{2}$, 
D.~Rowntree$^{3}$, A.J.~Sarty$^{8}$, H.~Schmieden$^{2}$, 
S.~\u{S}irca$^{10}$, J.A.~Templon$^{9}$, M.~Thompson$^{1}$, A.~Wagner$^{2}$, 
Th.~Walcher$^2$, M.~Weis$^{2}$, J.~Zhao$^{3}$, Z.~Zhou$^{3}$}}
\end{center}
\noindent
{\em \footnotesize{1 -- School of Physics, The University of Melbourne,
    Parkville 3052, VIC, Australia; \\
2 -- Institut f\"{u}r Kernphysik, Universit\"{a}t Mainz, D-55099 Mainz, Germany;\\
3 -- Laboratory for Nuclear Science, MIT, Cambridge, MA 02139, USA;\\
4 -- TJNAF, Newport News, VA, USA;\\
5 -- Department of Physics and Astr., California St. U., Los Angeles, CA 90032, USA;\\
6 -- Department of Physics and Astr., U. of Kentucky, Lexington, KY 40506, USA;\\
7 -- Institut f\"{u}r Kernphysik, Technische U. Darmstadt, D-64289 Darmstadt, Germany;\\
8 -- Department of Physics, Florida State University, Tallahassee, FL 32306,
USA;\\
9 -- Department of Physics and Astronomy, U. of Georgia, Athens,GA 30602, USA;\\
10 -- Institute "Jo\u{s}ef Stefan", University of Ljubljana, SI-1001 Ljubljana,
Slovenija;}
}\\ 
\\
A high-resolution study of the $(e,e'p)$ reaction on $^{4}\!He$ was 
carried out at the Institut f\"{u}r Kernphysik in Mainz, Germany.
The high quality 100~\% duty factor electron beam, and the
high-resolution three-spectrometer-system of the A1 collaboration were used.
The measurements were done in parallel kinematics at a central
momentum transfer $|\vec{q}|$~=~685~MeV/c, and at a central energy
transfer $\omega$~=~334~MeV, corresponding to a value of the 
$y$-scaling variable of +140~MeV/c. In order to enable the Rosenbluth
separation of the longitudinal $\sigma_{L}$ and transverse $\sigma_{T}$ 
response functions (as defined in [1]), three measurements at different 
incident beam energies,
corresponding to three values of the virtual photon polarization $\epsilon$,
were performed.\\
The absolute $(e,e'p)$ cross section for 
$^{4}\!He$ was obtained as a function of missing energy and missing momentum.
A distorted spectral function, $S^{dist}(E_{m},p_{m})$ was
extracted from the data using the $cc1$ prescription for the elementary
off-shell e-p cross section (see ref.[2]), where
 \be
S^{dist}(p_{m},E_{m})=\frac{1}{p^{2}_{p}\sigma_{ep}}
\frac{d^{6}\sigma}{d\Omega_{e}d\Omega_{p} dp_{e}dp_{p}}
\ee
For the two-body breakup channel the experimental results were 
compared to the theoretical calculations performed by Schiavilla {\em et al}. 
[3] and Forest {\em et al}. [4], and to the earlier 
experimental momentum distributions measured at NIKHEF by van den Brand {\em
  et al}. [5] and the new results from MAMI by Florizone [6] as shown in 
Fig. \ref{fig1}. 
For the continuum channel, recent calculations for the $^{4}\!He$ spectral
function by Efros {\em et al}. [7] are presented in Fig. 2,B to compare
with the experimental results.\\
A Rosenbluth separation was performed for both the two-body
breakup and for continium channels, and preliminary results were obtained.
The ratio $\sigma_{L}/\sigma_{T}$ was determined 
and compared with predictions (see Figure \ref{fig2}).\\
The measurements show no significant strength corresponding to the 
$(e,e'p)$ reaction channel for missing energy values 
$E_{m}\,\geq \,50\, MeV$ as shown in Fig. \ref{fig2}.\\ \\
{\bf References}\\
1. S.Boffi {\em et al}., Nucl. Phys. {\bf A435} (1985) 697\\
2. T. de Forest, Nucl. Phys., {\bf A392}, 232 (1983)\\
3. R. Schiavilla {\em et al}., Nucl. Phys. {\bf A449} 219 (1986)\\
4. J.L. Forest {\em et al}., Phys. Rev {\bf C54}, 646 (1996)\\
5. J.F.J. van den Brand {\em et al}., Nucl. Phys. {\bf A534} (1991) 637\\
6. R.E.J. Florizone {\em et al}., to be published in Phys. Rev. {\bf C}\\
7. V. D. Efros, W. Leidemann, G. Orlandini, Phys. Rev. {\bf C58} (1998) 582\\
\end{document}